\documentstyle[draft,aps,prl,twocolumn,epsf]{revtex}
\begin{document}
\title{ Incompleteness of Representation Theory:
Hidden symmetries and Quantum Non-Integrability}

\author{Dimitri Kusnezov
     \footnote{E--mail: dimitri@nst.physics.yale.edu} }
\address{ Center for Theoretical Physics,
              Sloane Physics Laboratory,
              Yale University, New Haven, CT 06520-8120 \\}

\date{December 12, 1996}

\maketitle

\vspace{3cm}

\begin{abstract}
Representation theory is shown to be incomplete in terms of enumerating all
integrable limits of quantum systems. As a consequence, one can find exactly
solvable Hamiltonians which have {\it apparently} strongly broken symmetry. The
number of these hidden symmetries depends upon the realization of the
Hamiltonian.
\end{abstract}

\draft

%

\narrowtext

When regular regions are discovered in the parameter space of a system, where
the parameter might be an external field applied to the Hydrogen atom, it
usually indicates the existence of new (approximate) integrals of motion, and
consequently quantum numbers\cite{kepler}. In studies of many--body systems,
the large number of degrees of freedom generally precludes one from using 
methods introduced in simpler one and two dimensional problems. However, one
often starts with a group theoretical formalism, and while classical  analyses
stop being practical, group theory can readily identify exactly solvable limits
of such systems, and correspondingly, quantum numbers.  These exactly solvable
limits are referred to as dynamical symmetries since they arise from the nature
of the interactions. It has long been assumed that there is a precise relation
between the exactly solvable limits or integrability of Hamiltonians based on
some Lie algebra $ {\cal G}$\cite{Zhang}, and the dynamical symmetries obtained
from representation theory\cite{fi}. Indeed there is now a large literature on
studies relating classical chaos to breaking of dynamical symmetries, its
consequences in random matrix theory, as well as relations to exactly solvable
systems\cite{barut}. It is interesting to consider now, whether group theory as
it stands actually identifies all integrable limits, or whether there are
hidden symmetries lingering in the parameter space of interactions.

A common starting point of group theoretical analyses is the identification of 
a dynamical algebra ${\cal G}$ for a given quantum system.  There is such an
algebra when a Hamiltonian $H$ can be expressed in terms of  the generators of
${\cal G}$. (For this study, we will focus on models based on real forms of
simple and semi-simple classical Lie algebras, denoted by ${\cal G}$.) The next
step is to use the techniques of representation theory to identify all
subalgebra embeddings, or group chains, consistent with the problem at hand. If
there are $n$ such group chains, one arrives at a decomposition of the  form:
\begin{equation}
{\cal G}\supset 
\begin{array}{lcccc}
{\cal G}_{11} & \supset  & {\cal G}_{12} & \supset  & \cdots  \\ 
&  & \vdots  &  &  \\ 
{\cal G}_{n1} & \supset  & {\cal G}_{n2} & \supset  & \cdots 
\end{array}
.
\end{equation}
Associated with each of the algebras are Casimir invariants $C({\cal G}_{kl})$.
All the invariants along a particular group chain (row) are in involution, and
these form (together with invariants associated with missing labels) a complete
set of constants of the motion. When the Hamiltonian can be expressed in terms
of the invariants along a single chain, then the system is said to have a
dynamical symmetry, and the problem is exactly solvable. Generally, it is found
that the linear and quadratic Casimir invariants from all the limits in (1)
provide a complete basis for a Hamiltonian written in terms of one and two body
interactions. These arguments, and the studies up to now, have hinged on the
belief that the group chains obtained from representation theory uniquely
enumerate all exactly solvable forms of the Hamiltonian. We will see that a
revision of this assertion is necessary.

Each of the dynamical symmetry limits characterized in Eq. (1) can be obtained
from Dynkin's theory of the embedding of subalgebras\cite{dynkin}. However,
they are only defined up to inner automorphisms. That is to say, each embedded
subalgebra is only a representative element of an equivalence class of
subalgebras, each with the same set of induced quantum numbers. We will show
here that while representation theory does provide all of the embeddings, it
does not provide all the dynamical symmetries. The inner automorphisms, which
are neglected, can provide `hidden' symmetries. One can have Hamiltonians which
are linear combinations of invariants from several or even all of the different
group chains in (1), which nevertheless is still exactly solvable. In this
sense, {\sl representation theory is incomplete in terms of defining all
dynamical symmetries of a system}. Specific examples will be shown for
Hamiltonians which would be argued to be non-integrable, but which in fact are
exactly solvable (or equivalently, integrable). In general, the number of 
dynamical symmetries can be richer than previously thought.

Consider a Hamiltonian $H$ which can be written in terms of generators of a
Lie algebra ${\cal G}$. We consider first the {\it automorphisms} of the
algebra ${\cal G}$. These are one-to-one mappings ${\cal T}$ of the
generators onto themselves, which naturally preserves the commutation
relations: ${\cal T}:{\cal G}\rightarrow {\cal G}$. There are many well
known automorphisms, such as the elements of the Weyl group ${\cal W}$,
which generate rotations in the root space of the algebra, or the complex
conjugation operation, which transforms generators $X_k$ to $-X_k^{*}$,
generating the contragredient representations. Non-trivial automorphisms can
be readily identified from the symmetries of the Dynkin diagrams\cite{dynkin}. 
Of the various automorphisms, only a certain class
of seemingly irrelevant ones are important here.

Let ${\sf g}$ be the Lie group associated with the Lie algebra ${\cal G}$ of
a Hamiltonian $H$. There are two classes of automorphisms of ${\cal G}$:
inner and outer. The inner automorphisms are transformations of the elements
of ${\cal G}$ generated by a fixed member $g$ of the group ${\sf g}$. They
are of the form: 
\begin{equation}
X\rightarrow gXg^{-1},\qquad \qquad \mbox{\rm all }X\in {\cal G},
\end{equation}
where $g\in {\sf g}$ is held fixed and all generators $X$ are transformed.
Outer automorphisms are by definition all automorphisms not of this type.
Consequently, embeddings of subalgebras ${\cal G}^{\prime }\subset {\cal G}$
fall into two categories: those considered equivalent (or conjugate) and
those considered inequivalent (or non-conjugate). Two embeddings of ${\cal G}
^{\prime }\subset {\cal G}$ are said to be conjugate if they are related to
each other through an inner automorphism of ${\cal G}$. Otherwise the
embeddings are non-conjugate, and are related by outer automorphisms of $
{\cal G}$. Non-conjugate embeddings are important because the embeddings are
distinct (and consequently the quantum numbers associated with the
subalgebra as well), and they are (essentially) all identified from Dynkin's
theory for the classification of subalgebras\cite{dynkin}. What is not
classified are the inner automorphisms. The equivalence class of conjugate
embeddings are generally neglected, since there are an infinite number of
these and any member of this class provides an equivalent result from the
point of view of representation theory: the embeddings generated are
isomorphic, and the induced representations are identical. Interestingly,
this is not the case when one considers dynamical symmetries of 
the Hamiltonian. 

To understand the origins of these conjugate embeddings, it is convenient (and
still completely general) to consider bosonic or fermionic realizations of
${\cal G}$. We will use bosons, although all our results hold equally well for
fermions. Denoting the boson creation and annihilation operators as  $b_{\ell
\mu }^{\dagger }$, $b_{\ell \mu }$, the generators of a Lie algebra can be
expressed in terms of bilinears $b_{\ell ,\mu }^{\dagger }b_{\ell ^{\prime
},\mu ^{\prime }}$. The boson carries angular momentum $\ell $ and projection
$\mu $, where $\mu =-\ell ,...,\ell $, which allows for the construction of
scalar, Hermitian Hamiltonians.  (This implicitly requires the explicit
construction of the $O(3)$ algebra, which then appears in all the chains in Eq.
(1).) The inner automorphisms we explore are generated by gauge rotations of
one type of boson with respect to the others. In order to preserve their
spherical tensor character, one must transform all components of the boson
simultaneously: $b_{\ell ,\mu }\rightarrow e^{i\phi _\ell }b_{\ell ,\mu }$.
However, each type of boson can be separately transformed. In the phase space
representation of the boson operators,$b_{\ell ,\mu }=(q_{\ell ,\mu }+ip_{\ell
,\mu })/\sqrt{2}$, this can be viewed as inducing certain canonical
transformations. The effect of some of these transformations will be to
generate new families of exactly solvable Hamiltonians. This type of
automorphism will not effect the R--subalgebras (in the language of
Dynkin\cite{dynkin})  such as $U(n)\supset U(n-1)$, $O(n)\supset O(n-1)$ and
$Sp(2n)\supset Sp(2(n-1))$, but can appear only for the S-subalgebras ${\cal
G}^{\prime }\subset {\cal G}$ , such as $U(n)\supset O(n)$, and so forth. The
main distinction is that S-subalgebra embeddings are formed from linear
combinations of generators of ${\cal G}$ where as R-subalgebras are not.
Another way to view the conjugate embeddings is in the Cartan-Weyl basis
$\{h_i,e_\alpha \}$. For a given root $\beta $, if we perform the trivial
rescaling $e_\beta \rightarrow \mu e_\beta $, $e_{-\beta }\rightarrow \mu
^{-1}e_{-\beta }$, with $|\mu |=1$, then the commutation relations of ${\cal
G}$ are preserved, including those of the embedded subalgebra ${\cal G}^{\prime
}$. (This acts as the identity operation in the Weyl group.) However, this
inner automorphism is precisely the gauge rotation discussed above. The effect
is to take generators of a subalgebra which are linear combinations of the
generators of ${\cal G}$, and alter their form, by changing the relative
phases. It is for this reason that only S-subalgebras are effected.
As a consequence the new generators of ${\cal G}^{\prime }$ do not commute with
the old ones. Because we often have additional physical requirements on the
Hamiltonian, not all of these automorphisms are allowed (e.g. the $O(3)$
generators must remain invariant, since physical angular momentum should not be
changed).  Thus necessary (albeit not sufficient) conditions for the existence
of conjugate subalgebras which results in new dynamical symmetry limits 
are $(i)$ ${\cal G}$ must be generated by at least two types of
bosons (or fermions) and $(ii)$ ${\cal G}^{\prime }$ must be an S--subalgebra.

To exemplify this, consider two models. The first is the Vibron model\cite
{fi}, which is based on ${\cal G}=U(4)$ and describes collective excitations
in molecules. The basic ingredients are the scalar, $\sigma^{\dagger }$ $
(\ell^\pi =0^{+}),$ and vector $\pi_\mu ^{\dagger }(\ell^\pi =1^{-},\mu
=0,\pm 1)$ boson creation operators, and the corresponding annihilation
operators $\sigma$, $\widetilde{\pi}_\mu =(-)^\mu \pi_{-\mu }$. The
generators are created in the usual way from all bilinears $\sigma^{\dagger
}\sigma$, $\sigma^{\dagger }\widetilde{\pi}_\mu $, $\pi_\mu ^{\dagger }\sigma
$, $\pi_\mu ^{\dagger } \widetilde{\pi}_\nu $. Under the constraint that
physical dynamical symmetries contain the angular momentum algebra $O(3)$,
representation theory gives two dynamical symmetries\cite{fi}:

\begin{equation}
U(4)\supset 
\begin{array}{c}
U(3) \\
\\
O(4) \end{array} \supset  O(3) \supset  O(2),
\end{equation}
which correspond to non-rigid (U(3)) and rovibrator (O(4)) molecules. The
$U(3)$ R--subalgebra is generated only by $\pi $ bosons, and hence does not
contain any relevant inner automorphisms. The $O(4)$ S--subalgebra  satisfies
our criteria and indeed there are two conjugate embeddings of $U(4)\supset
O(4)$, related by the transformation $\pi _\mu \rightarrow -i\pi _\mu $,
$\sigma \rightarrow \sigma $. Define $L_\mu =\sqrt{2}[\pi ^{\dagger }\tilde \pi
]_\mu ^{(1)}$ $D_\mu =[\pi ^{\dagger }\sigma +\sigma ^{\dagger }\tilde \pi
]_\mu ^{(1)}$, and $D_\mu ^{\prime }=i[\pi ^{\dagger }\sigma -\sigma ^{\dagger
}\tilde \pi ]_\mu ^{(1)}$, where $\mu =\pm 1,0$ (the square brackets represent
the usual angular momentum coupling of spherical tensors). Then the two
conjugate embeddings, denoted $O(4)$ and $\overline{ O(4)}$, are obtained by
the six generators $\left\{ L_\mu ,D_\upsilon \right\} $ and $\left\{ L_\mu
,D_\upsilon ^{\prime }\right\} $\cite{onno}. Both contain the same $O(3)$
subalgebra as the $U(3)$ chain, generated by  $L_\mu $. Their respective Cartan
subalgebras are generated by $L_0,D_0$ and $ L_0,D_0^{\prime }$. Since $\left[
D_0,D_0^{\prime }\right] \neq 0$, the two algebras are not related by Weyl
reflections. However, they are related by a similarity transformation in $U(4)$
(but not in $O(4)$) of the form (2).  The full classification (3) from the
dynamical symmetry point of view should include $(iii)$ $ U(4)\supset
\overline{O(4)}\supset O(3)\supset O(2)$.

As another illustration, consider the Interacting Boson Model (IBM)\cite{fi}
, which describes collective excitations in nuclei. The basic ingredients
are $s^{\dagger }$ and $d_\mu ^{\dagger }$ ($\mu =\pm 2,\pm 1,0$) bosons,
which represent nucleons paired to $L=0$ similar to a cooper pair, and to $
L=2$ describing two-nucleon (or hole) excitations, respectively. The
dynamical group is $U(6)$ which is built from the 36 bilinears $s^{\dagger
}s,s^{\dagger }d_\mu ,d_\mu ^{\dagger }s,d_\mu ^{\dagger }d_\nu $, and
representation theory provides three well known dynamical symmetries: (i) $
U(6)\supset O(6)\supset O(5)\supset O(3)\supset O(2)$, (ii) $U(6)\supset
U(5)\supset O(5)\supset O(3)\supset O(2)$, and (iii) $U(6)\supset
SU(3)\supset O(3)\supset O(2)$. The most general expansion of the
Hamiltonian into one- and two-body interactions, which is Hermitian,
time-reversal invariant and scalar, can be expanded into six terms
consisting of the linear and quadratic Casimir operators from all three
limits\cite{fi}. Using our requirements, we find that there are two 
conjugate embeddings, denotes $\overline{
O(6)}$ and $\overline{SU(3)}$. While the generators of these limits have
been explored in the past\cite{piet,fi}, new dynamical symmetries were not
associated with them. Consider first $O(6)$, which is
generated by the 15 operators $\{L_\mu ,U_\mu ,Q_\mu \}$, where $L_\mu =
\sqrt{10}[d^{\dagger }\tilde d]_\mu ^{(1)}$ is the angular momentum, $U_\mu
=[d^{\dagger }\tilde d]_\mu ^{(3)}$ and $Q_\mu =[s^{\dagger }\tilde d_\mu
+d_\mu ^{\dagger }s]$. The conjugate $O(6)$, denoted $\overline{O(6)}$, 
is generated by $\{L_\mu ,U_\mu ,\overline{Q_\mu }\}$, where the
difference is in the 5 generators $\overline{Q_\mu }=i[s^{\dagger }\tilde d
_\mu -d_\mu ^{\dagger }s]$. As with $O(4)$ and $\overline{O(4)}$, the
invariants of the two conjugate $O(6)$ algebras do not commute, nor do their
Cartan subalgebras.

Unlike $U(4)$, the algebra of $U(6)$ admits another conjugate subalgebra. This
is the $U(6)\supset SU(3)$ embedding. $SU(3)$ can be generated by the
quadrupole operator $Q_\mu ^{\pm }=[s^{\dagger }\tilde d_\mu +d_\mu ^{\dagger
}s]\pm (\sqrt{7}/2)[d^{\dagger }\tilde d]_\mu ^{(2)}$ and the angular momentum
$L_\mu $, using either $\{Q_\mu ^{-},L_\nu \}$ or $\{Q_\mu ^{+},L_\nu
\}$\cite{fi}. The Cartan generators of these $SU(3)$ algebras are 
$\{L_0,Q_0^{-}\}$ or $\{L_0,Q_0^{+}\}$, which do not commute but are related
through an inner automorphism. Physically this inner automorphism transforms
the Hamiltonian of a prolate nucleus into one for an oblate nucleus, or
equivalently, it can be related to particle-hole conjugation. Thus the complete
dynamical symmetry classification of the $U(6)$ model includes the two
additional ones $(iv)$ $ U(6)\supset \overline{SU(3)}\supset O(3)\supset O(2)$
and $(v)$ $U(6)\supset\overline{O(6)}\supset O(5)\supset O(3)\supset O(2)$.

Consider now the consequences of these conjugate subalgebras. A conjugate
dynamical symmetry Hamiltonian, written in terms of the invariants $\{
\overline{C}\}$ of a conjugate subalgebra group chain, can be re--expressed
as an expansion into some or all of the invariants $\{C\}$ predicted from
representation theory, Eq. (1): 
\begin{equation}
H=f(\{\overline{C}\})=\sum_{ijk}\alpha _{ijk}g_i(C({\cal G}_{jk})).
\end{equation}
The right side of (4) is a general expansion into the Casimir invariants of
various (or all) group chains in (1). The left hand side is an exact
dynamical symmetry. Again, this type of mapping is not necessarily outside
the range of physical interest. In the examples discussed above, we can now
realize additional dynamical symmetries. If we defined the $\overline{O(4)}$
Hamiltonian in terms of the quadratic invariants (denoted $C_2$) of that
chain, $H(\overline{O4})=\alpha C_2[\overline{O(4)}]+\beta C_2[O(3)]$,  we
have: 
\begin{eqnarray}
H(\overline{O4}) &=&-\alpha C_2[O(4)]+4\alpha (N-1)C_1[U(3)] \\
&&-4\alpha C_2[U(3)]+(2\alpha +\beta )C_2[O(3)]+6\alpha N.  \nonumber
\end{eqnarray}
This relation (and (6),(7) below) can be obtained by expanding $H$ into the
generators of ${\cal G}$ and then re-expressing them in the Casimir invariants 
of the other limits.  
Here $C_1[U(3)]=\hat n_\pi =\pi ^{\dagger }\cdot \tilde \pi $ and $C_2[U(3)]=
\hat n_\pi ^2$ are the linear and quadratic Casimir operators for $U(3)$, and
$C_2[O(3)]=\hat L^2$ is the quadratic Casimir of $O(3)$. While this strongly
mixes the two group chains given from representation theory in (3), the inner
automorphism allows for this new parametric family of exactly solvable
Hamiltonians, in this case with an $O(4)$ spectrum. In fact it has the same
spectrum as $H=\alpha C_2[O(4)]+\beta C_2[O(3)]$. A consequence
is that one can view the equivalence of these two Hamiltonians as a parameter
symmetry  in the space in one- and two-body interactions. This  conjugate
$O(4)$ Casimir have been recently examined, including such a parameter
transformation\cite{shir}.

In the IBM, we can take the general forms of the $\overline{O(6)}$ and $
\overline{SU(3)}$ dynamical symmetry Hamiltonians, which we write as $H(
\overline{O(6)})=\alpha P^{^{\prime }\dagger }P^{^{\prime }}+\beta U\cdot
U+\gamma L\cdot L$ and $H(\overline{SU(3)})=\alpha Q^{+}\cdot Q^{+}+\beta
L\cdot L$, and obtain the two families of exactly solvable Hamiltonians: 
\begin{eqnarray}
H(\overline{O(6)}) &=&-\alpha P^{\dagger }P+(\beta -4\alpha )U\cdot U \\
&&+(\gamma -\frac 25\alpha )L\cdot L+4\alpha (2-N)\widehat{n}_d  \nonumber \\
&&+4\alpha \widehat{n}_d^2+2N(N-1).  \nonumber \\
H(\overline{SU(3)}) &=&5\alpha \widehat{n}_d+\alpha \widehat{n}_d^2-\alpha
Q^{-}\cdot Q^{-}-2\alpha P^{\dagger }P \\
&&-6\alpha U\cdot U+(\beta -\frac 7{20}\alpha )L\cdot L  \nonumber \\
&&+2N(N+4)  \nonumber
\end{eqnarray}
(In the $O(6)$ limits, $C_2[O(6)]$ is related to the pairing operator $P$
\cite{fi}. In the usual $O(6)$ limit, $P=ss-\tilde d\cdot \tilde d$, while
in $\overline{O(6)}$ it is $P^{\prime }=ss+\tilde d\cdot \tilde d$.) Eq. (6)
involves Casimir operators from both the $O(6)$ and $U(5)$ limits, and is
known to be classically integrable, due to the common $O(5)$ subalgebra\cite
{Zhang}. But our statement here is stronger than that. We see here that this
particular combination of operators is not only integrable, but is
an $O(6)$ dynamical symmetry. Eq. (7) would have been termed
non-integrable, because it appears to be a case of strongly broken $SU(3)$, as
it includes Casimir invariants from all three distinct dynamical symmetries
of the IBM. However, it is clearly exactly solvable, and is diagonal in an 
$SU(3)$ basis.

In terms of the gauge rotations of the bosons, for the conjugate $SU(3)$
algebras, $\phi _2=0,\pi $ ($\pi /2,3\pi /2$ are not allowed by
time-reversal invariance), while for the $O(6)$ ($O(4)$) algebra, $\phi
_2(\phi _1)=0,3\pi /2$ ($\pi ,\pi /2$ being equivalent). Other values of $
\phi $ are not allowed due to the constraints that the Hamiltonians are
time-reversal invariant. Generally, the number of conjugate embeddings 
depends upon the system of interest. For
example, in the IBM, the $SU(3)\supset O(3)$ embeddings is unique due to the
constraint that the Hamiltonian is scalar. However, in the molecular $SU(3)$
model of Ref. \cite{oss1} for planar rotations and vibrations, there are two 
$O(3)$ embeddings according to our criteria. In other models, such as the
proton-neutron IBM, there are even more examples since one can perform
independent rotations of neutron and proton boson operators, resulting in
many additional dynamical symmetries, or in the study of octupole deformed
nuclei where one can transform $s,p,d$ and $f$ bosons\cite{oct}. Similarly,
this will also be possible in the extensions of the vibron model to
polyatomic molecules, which is based in $U(4)\otimes \cdots \otimes U(4)$
\cite{ch}. In studies of chaos in triatomic molecules using $U(4)\otimes U(4)
$, it is the conjugate embedding of $\overline{U(4)}$ which explains the
recently observed regularity in the parameter space\cite{ch}. 

We have shown that the number of group chains provided by representation theory
does not necessarily correspond to the number of dynamical symmetries of a
system. We have further seen that (apparent) breaking of  dynamical symmetries
does not guarantee non-integrability. These hidden symmetries generate new
families of dynamical symmetry Hamiltonians which contain generators from some
or all of the different group chains in Eq. (1).  They can also have different
physical properties, as in the case of $SU(3)$ and $\overline{SU(3)}$ in the
IBM. The former corresponds to the spectra of oblate deformed nuclei, with 
negative quadrupole moment; the latter to prolate nuclei with positive
quadrupole moment - these are not only physically distinct, but both cases are
needed for a complete physical picture. In the studies of chaos and
quantum-nonintegrability, one must clearly consider conjugate embeddings which
will generally suppress chaos in regions in the parameter space which would
have previously been expected to be chaotic. In general there are an infinite
number of inner automorphisms - however only a finite number may be consistent
with physics (in the above examples, these are scalar, Hermitian Hamiltonians
with time reversal invariance). In nuclei, it is evident that strongly 
deformed or $\gamma -$soft nuclei are not limited to Hamiltonians in one of the
group chains. In principle, all conjugate embeddings can be enumerated. It
would be interesting to see whether one can develop within representation
theory, a general prescription to classify the non-trivial inner automorphisms,
and to explore supersymmetric automorphisms which relate fermions to bosons.

I would like to thank A. Dieperink, K. Heyde, J. Ginocchio, F. Iachello,  P.
Van Isacker and A. Leviatan for useful discussions, and the ECT$^*$ for their
support. This work was supported by DOE grant DE-FG02-91ER40608.


\begin{references}

\bibitem{kepler} D. Delande and J.C. Gay, {\sl Phys. Rev. Lett.} {\bf 59}
(1987) 1809.

\bibitem{Zhang}  W.M.Zhang and D.H.Feng, {\sl Phys. Rep.} {\bf 252} (1995)
1; W.M.Zhang, C. Martens, D.H.Feng and J.M.Yuan, {\sl Phys. Rev. Lett.} {\bf 
61} (1988) 2167; W.M.Zhang and D.H.Feng, {\sl Phys. Rev.} {\bf C43} (1991)
1127; Y. Alhassid, A. Novoselsky and N.Whelan, {\sl Phys. Rev. Lett.} {\bf 65
} (1990) 2971; Y. Alhassid and N.Whelan, {\sl Phys. Rev. Lett.} {\bf 67}
(1991) 816.

\bibitem{fi}  See for example: F. Iachello and A. Arima, {\sl The
Interacting Boson Model} (Cambridge Press, Cambridge, 1987); F. Iachello and
R. Levine, {\it Algebraic Theory of Molecules} (Oxford Press, Oxford, 1995).

\bibitem{barut}  {\sl Dynamical Groups and Spectrum
Generating Algebras}, Eds. A.Barut, A. Bohm and Y.Ne'eman (World Scientific,
Singapore, 1987); {\sl Few Body Methods, Principles and Applications}, Eds.
T.K.Lim, C.G.Bao, D.P.Hou and J.S.Huber (World Scientific, Singapore, 1986); 
{\sl \ Symmetries in Science V}, Eds. B. Gruber, L.C. Biedenharn, and H.D.
Doebner, (Plenum, New York, 1991).

\bibitem{dynkin}  E. Dynkin, {\sl Mat. Sb.} {\bf 30} (1952) 349; {\sl Amer.
Math. Soc. Trans. [2]}, {\bf 6} (1957) 111; {\sl Tr. Mosk. Mat. Obsc.} {\bf 
1 } (1952) 39; {\sl Amer. Math. Soc. Trans. [2]}, {\bf 6} (1957) 245.



\bibitem{onno}  O.van Roosmalen, Ph.D. thesis, (1982) unpublished.

\bibitem{piet}  P. Van Isacker, A. Frank and J.Dukelsky, {\sl Phys. Rev.} 
{\bf C31} (1985) 671.

\bibitem{shir}  A.M.Shirokov and N.A.Smirnova, Moscow State University
preprint (1996).

\bibitem{oss1}  F. Iachello and S. Oss, {\sl J. Chem. Phys.} {\bf 104}
(1996) 1.

\bibitem{oct}  D.Kusnezov, {\sl J. Phys. A: Math. and General} {\bf 23},
5673 (1990); {\sl J. Phys. A: Math. and General} {\bf 22}, 4271 (1989).

\bibitem{ch}  J.M. Champion, Ph.D. thesis, Univ. J.Fourier, Grenoble (1996).
\end{references}
\end{document}